\documentclass[conference]{IEEEtran}
\usepackage{placeins}
\usepackage{multirow}
\usepackage{subcaption}
\usepackage{booktabs}
\usepackage{verbatim}
\usepackage{hyperref}
\usepackage{cite}
\usepackage{xcolor}
\usepackage{soul}
\usepackage{graphicx}
\usepackage{amsmath}
\usepackage{float}
\usepackage{balance}
\usepackage{subcaption} %  for subfigures 
\usepackage{algorithm}
\usepackage{algpseudocode}
\usepackage{bm}
\newcommand{\sph}{\mathsf{SPHINCS^+}}
\newcommand{\kb}{\mathsf{Kyber}}
\newcommand{\dil}{\mathsf{Dilithium}}

\def\BibTeX{{\rm B\kern-.05em{\sc i\kern-.025em b}\kern-.08em
    T\kern-.1667em\lower.7ex\hbox{E}\kern-.125emX}}
    
\begin{document}

\title{OptHQC: Optimize HQC for High-Performance Post-Quantum Cryptography}

\author{
    \IEEEauthorblockN{ Ben Dong, Hui Feng, Qian Wang}
    \IEEEauthorblockA{Department of Electrical Engineering, University of California, Merced, CA, USA}
    \IEEEauthorblockA{cdong12@ucmerced.edu, hfeng9@ucmerced.edu, qianwang@ucmerced.edu}
}

\maketitle

\begin{abstract}
As post-quantum cryptography (PQC) becomes increasingly critical for securing future communication systems, the performance overhead introduced by quantum-resistant algorithms presents a major computing challenge. HQC (Hamming Quasi-Cyclic) is a newly standardized code-based PQC scheme designed to replace classical key exchange methods. In this paper, we propose OptHQC, an optimized implementation of the HQC scheme to deliver high-performance cryptographic operations. Our approach provides a comprehensive analysis of each computational blocks in HQC and introduces optimizations across all three stages: key generation, encryption, and decryption.
We first exploit data-level sparsity in vector multiplication to accelerate polynomial operations during vector generation.
We then leverage instruction-level acceleration (e.g., AVX2) in hash computation to further improve performance.
Last, we transform multiplication into lookup table indexing and optimize memory access patterns in syndrome computation and error vector recovery, which are the most computationally intensive operations in HQC.
Overall, OptHQC achieves an average 55\% speedup over the reference HQC implementation on CPU.

%Besides, we also introduce SIMD-friendly memory layouts to improve throughput. 
\end{abstract}

\vspace{-0.25cm}
\section{Introduction}

With the computational superiority of the quantum computer, the currently deployed cryptographic algorithms based on the integer factorization problem (e.g., RSA), the discrete logarithm problem (e.g., DH), or the elliptic-curve discrete logarithm problem (e.g., ECC) will all be broken in polynomial time. Therefore, it is crucial to update and upgrade the current algorithms to withstand quantum attacks. The National Institute of Standards and Technology (NIST) has initiated the standardization process to promote newly designed post-quantum cryptographic algorithms \cite{moody2021nist}.  
The selected PQC schemes include CRYSTALS-$\kb$ \cite{bos2018crystals}, CRYSTALS-$\dil$ \cite{ducas2018crystals}, and Falcon \cite{fouque2018falcon}, which are lattice-based, and $\sph$ \cite{bernstein2019sphincs+}, which is hash-based. Most recently, NIST finalized HQC as the only code-based scheme in the PQC standards \cite{nist2025hqc}.

%Among the selected schemes, CRYSTALS-$\kb$ \cite{bos2018crystals}, CRYSTALS-$\dil$  CRYSTALS-$\dil$ \cite{ducas2018crystals}, and $\fal$ \cite{fouque2018falcon} are lattice-based. and $\sph$ \cite{bernstein2019sphincs+} are hash-based. Most recently, NIST announced the HQC, the only code-based scheme in the standards \cite{nist2025hqc}.

While HQC diversifies the KEM portfolio, its computational cost is dominated by several compute-intensive components:
(i) the concatenated decoding stage, where a byte-parallel Reed–Muller decoder feeds into a Reed–Solomon decoder, which largely determines the decoding latency;
(ii) the SHAKE-based seed expansion and domain-separated hashing operations used for randomness generation and key derivation; and
(iii) constant-time dense-by-sparse vector and polynomial multiplications, thereby exerting significant pressure on memory bandwidth. Prior evaluation shows that HQC delivered only \(\sim\!2\%\) of Kyber’s throughput under identical conditions ~\cite{epquic}. Our profiling results on HQC, shown as in Figure \ref{fig:hqc_perf_bar} confirm these three hotspots as the dominant contributors to runtime, motivating the proposed optimizations to reduce overall latency.

%In the category of mathematical problems, $\kb$, $\dil$, and $\fal$ are lattice-based algorithms, while $\sph$ is based on the cryptographic hash functions. The most recently added $\hq$ are categorized as a code-based scheme. In our previous investigation, the algorithm achieved about 2\% of the performance when compared to the Kyber KEM algorithm. After our investigation, we have identified the computational operation performed at each step of the algorithm and used the information to identify the bottleneck. After the bottleneck has been identified, we have implemented optimization into the algorithm.

The primary contributions of our paper are as follows:
\begin{itemize}
    \item We profile the performance of HQC by decomposing it into key computational stages, including hashing, sampling, encoding, syndrome computation, and concatenated Reed–Muller/Reed–Solomon decoding. We further identify performance overheads, quantify each block’s runtime contribution, and analyze the optimization within each stage.
    \item We implement lane‑interleaved Keccak (SHAKE) with unrolled rounds, aligned loads/stores, and absorb/squeeze fusion, and provide a minimal adapter to the HQC codebase. The kernel reduces SHAKE PRNG and seed‑expansion time by 2×. 
    \item We implemented an efficient sparse $\times$ dense vector algorithm, using the polynomial shifting technique, which resulted in roughly 22\% speed up at key generation, 60\% speed up at encryption, and 35\% speedup at decryption.
    \item We optimize the Reed--Solomon encode/decode process by replacing costly multiplications with precomputed lookup tables, converting them into constant-time, cache-efficient XOR and shift operations that also preserve side-channel resistance.

\end{itemize}
%The rest of the paper is organized as follows. The basics of the HQC algorithm are given in the next section. Section III summarizes the related work. In Section IV, we present the details of implementing the optmization. Section V provides a thorough performance analysis to evaluate the applicability of PQC applications. Section VI concludes the paper.

%\vspace{-0.1cm} % Add some vertical space between the tables
\section{Backgrounds} 
% \subsection{Overview of Post-quantum Crypto}
% PQC refers to cryptographic algorithms designed to be resistant to attacks facilitated by quantum computers. PQC encompasses various cryptographic families, including lattice-based, hash-based, code-based, multivariate, and isogeny schemes. Notably, hash-based and lattice-based schemes have been selected for standardization, considering factors such as performance, security, and applicability. Among those diverse proposed PQC algorithms, hash-based schemes are more mature than others as they have undergone extensive study years ago and are well understood in terms of their security properties. Lattice-based cryptography delves into the study of algebraic structures such as Ring Learning With Errors (Ring-LWE) or Short Integer Solution (SIS) of lattices. NIST's standardization process has recognized the efficiency of lattice-based cryptography for the selection of three lattice-based schemes for standardization. These include $\kb$ for the key-encapsulation mechanism (KEM), as well as $\dil$ and $\fal$, both being selected for signature generation. Most recently, one of the code-based schemes, HQC, was selected as an additional KEM algorithm to enhance diversity within the KEM family, which previously included only lattice-based candidates. However, HQC suffers from performance limitations.

\subsection{HQC Computation Flow}
HQC is a secure code-based KEM whose security relies on a decision version of syndrome decoding on structured QC codes. It combines (i) a random QC code with a public parity-check matrix for security, and (ii) employs a fixed concatenated code consisting of an external shortened Reed–Solomon code and an internal duplicated Reed–Muller code for error correction.

HQC employs SHAKE functions as a PRNG for sampling, generating either dense coefficients directly or sparse ones through a fixed-weight sampler. Domain-separated SHAKE-derived functions $G(\cdot),H(\cdot),K(\cdot)$ are used for seed/digest formation and key derivation.

HQC.\emph{Encrypt}~\ref{alg:hqc_encrypt} samples fixed-weight $e,r_1,r_2$, encodes $m$ with generator $G$, forms $u=r_1+h\cdot r_2$, and $v=mG+s\cdot r_2+e$, yielding the ciphertext as $c=(u,v)$. \emph{Decrypt}~\ref{alg:hqc_decrypt} computes $v-u\cdot s$ and applies Reed–Muller then Reed–Solomon decoding to recover $m'$. 

\vspace{-2mm}
\begin{algorithm}[H]
\caption{\textbf{HQC.PKE.Encrypt}}
\label{alg:hqc_encrypt}
\begin{algorithmic}
\State \textbf{Input:} Public key $pk = (h, s)$, message $m$
\State Sample $e \leftarrow \mathcal{R}$ with $\omega(e) = \omega_e$
\State Sample $r = (r_1, r_2) \leftarrow \mathcal{R}^2$ with $\omega(r_1) = \omega(r_2) = \omega_r$
\State $\bm{u \gets r_1 + h \cdot r_2}$
\State $\bm{v \gets mG + s \cdot r_2 + e}$
\State \Return $c = (u, v)$
\end{algorithmic}
\end{algorithm}
\vspace{-5mm}
\begin{algorithm}[H]
\caption{\textbf{HQC.PKE.Decrypt}}
\label{alg:hqc_decrypt}
\begin{algorithmic}
\State \textbf{Input:} Secret key $sk$, ciphertext $c = (u, v)$
\State \Return  $\bm{C.\mathrm{Decode}(v - u \cdot s)}$
\end{algorithmic}
\end{algorithm}
\vspace{-4mm}

HQC.KEM wraps the PKE using the HHK transform with implicit rejection, which consists of three subroutines as \emph{keyGen}, \emph{Encap}, and \emph{Decap}. \emph{KeyGen}~\ref{alg:hqc_keygen} samples seeds, regenerates $h$, draws fixed-weight secrets $x$ and $y$, computes $s = x + h \cdot y$, and embeds a secret $\sigma$ for use in \emph{Decap}. In \emph{Encap}, the sender derives a session key while generating a ciphertext (calling \emph{Encrypt} that encapsulates the shared secret. In \emph{Decap}, the receiver decrypts the ciphertext (calling \emph{Decrypt}), and then re-encrypts to verify correctness, and securely rejects invalid ciphertexts without leaking side-channel information.
We reference standard parameter sets (HQC-1, -3, and -5), as shown in Table \ref{HQC_para}, throughout our implementation and optimization results.

\vspace{-2mm}
\begin{algorithm}[H]
\caption{\textbf{HQC.KEM.KeyGen}}
\label{alg:hqc_keygen}
\begin{algorithmic}
\State \textbf{Input:} $\mathbf{param} = (n, k, \delta, \omega, \omega_r, \omega_e)$
\State Sample $h \leftarrow \mathcal{R}$
\State Sample $(x, y) \leftarrow \mathcal{R}^2$ with $\omega(x) = \omega(y) = \omega$
\State $sk = (x, y, \sigma)$
\State $\bm{pk = (h, s = x + h \cdot y)}$
\State \Return $(pk, sk)$
\end{algorithmic}
\end{algorithm}
\vspace{-4mm}

\begin{table}[h!]
\centering
\caption{HQC parameter sets}
\vspace{-2mm}
\begin{tabular}{|c|c|c|c|c|c|}
\hline
Level & $N$ & $\omega$ & $\omega_e = \omega_r$ & Reed-Solomon & Reed-Muller \\
\hline
HQC-1 & 17,669 & 66  & 75  & [46,16,31]  & [384,8,192] \\
HQC-3 & 35,851 & 100 & 114 & [56,24,33]  & [640,8,320] \\
HQC-5 & 57,637 & 131 & 149 & [90,32,59]  & [640,8,320] \\
\hline
\end{tabular}
\vspace{-0.5cm}
\label{HQC_para}
\end{table}

% \subsection{SIMD Instruction}
% instruction}
\subsection{Related Works}

The implementation of the HQC \cite{rasoptimizing} key encapsulation mechanism has been widely explored since its inclusion in the NIST PQC standardization process. Prior works can be grouped into four main categories: hardware, software, hardware/software co-design and side-channel–resilient implementations. Hardware-based studies focus on FPGA and ASIC designs that enhance performance and resource efficiency through pipelined polynomial multiplication, shared SHAKE modules dual-clock architectures \cite{roy2022fasthqc, deshpande2023fast}. Software implementations, on the other hand, emphasize algorithmic and data-structure optimizations on general-purpose processors or embedded CPUs, such as bit-slicing, loop unrolling and memory-efficient encoding for ARM Cortex-M4 \cite{kim2025optimized}. The hardware/software co-design approach balances speed and flexibility by offloading heavy computations to hardware accelerators while keeping control flow in software, as demonstrated in Code-Based Cryptography in IoT: A HW/SW Co-Design of HQC \cite{schoffel2022code}. Meanwhile, side-channel–resilient implementations aim to defend against timing, power and fault attacks using constant-time decoding, masked Keccak modules and verification-based decapsulation \cite{goy2022new}. In contrast, our proposed OptHQC framework targets high-performance, portable software optimization of the HQC scheme. It exploits thread-level and data-level parallelism together with vectorization to accelerate key generation, encryption and decryption, achieving significant speedups on general-purpose processors without requiring specialized hardware.

\section{Methodology}

\subsection{HQC Profiling Result}
\begin{figure*}[t]
    \centering
    \includegraphics[width=0.82\linewidth]{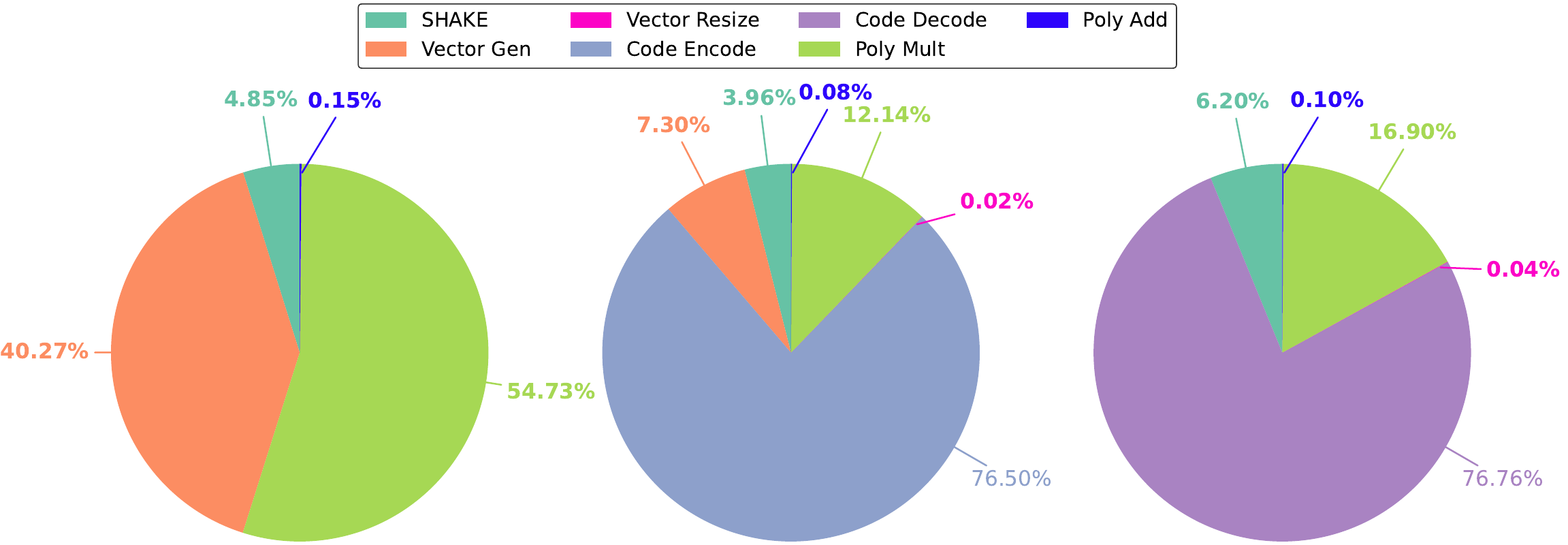}
    \caption{\small
    Percentage breakdown of computational costs in HQC-128. 
    The three pie charts illustrate the operation distribution in the \textit{Key Generation}, \textit{Encryption}, 
    and \textit{Decryption} algorithms. Poly Mult and Code Encode/Decode dominate the runtime, while 
    SHAKE, Vector Gen, Vector Resize, and Poly Add contribute relatively minor overhead.}
    \label{fig:hqc_piecharts}
    \vspace{-5mm}
\end{figure*}
Following the HQC specification, we begin by profiling the computational cost of the \texttt{KeyGen}, \texttt{Encrypt}, and \texttt{Decrypt} subroutines, shown in Figure \ref{fig:hqc_piecharts}. The HQC execution starts with SHAKE256-based randomness generation, which produces dense coefficients directly and samples fixed-weight sparse indices for secrets and error vectors.  Hash-related operations contribute roughly $5\%$ of the overall HQC runtime.

The vector generation phase dominates the computation in \texttt{KeyGen}, as it constructs the sparse vectors $(x, y)$ and the dense vector $h$ through sampling derived from the public seed. 
The following computation involves polynomial multiplications, usually dense-by-sparse products such as $h \cdot r_2$ and $s \cdot r_2$. All HQC wrapper operations, including \texttt{KeyGen}, \texttt{Encrypt}, and \texttt{Decrypt}, perform such polynomial multiplications. 
The encoder and decoder dominate the encryption and decryption processes. They apply a concatenated scheme combining an outer shortened Reed–Solomon (RS) code with an inner duplicated Reed–Muller (RM) code.

Based on our profiling results, we explore methods to optimize HQC performance on general computing platforms by leveraging accelerated polynomial multiplication (Section~\ref{sec:Mul}), optimized Keccak implementations (Section~\ref{sec:shake}), and efficient encoding and decoding (Section~\ref{sec:enc}) strategies.

\subsection{Multiplication Optimization in AVX}
\label{sec:Mul}

We implement an adaptive ring-multiplication path that switches between sparse–dense and dense–dense strategies based on the Hamming weight of the first operand. This targets directly at the \texttt{Poly Mult} section in the Figure~\ref{fig:hqc_piecharts}. The routine first counts nonzero coefficients in $a_1$; if the number is below a sparsity threshold, the computation routes to a sparse–dense optimization. Otherwise, it uses dense Toom–Cook multiplication followed by modular reduction.

For the sparse case, the product is synthesized as a superposition of rotated copies of the dense operand:
\begin{equation}
    \label{eq:multiply}
    a_1(X)\cdot a_2(X)\bmod (X^N-1)=\bigoplus_{i\in \operatorname{Supp}(a_1)} X^i\cdot a_2(X)
\end{equation}
Each rotation corresponds to a cyclic shift by a support index followed by XOR accumulation. This scales with the number of nonzero coefficients rather than polynomial degree, providing significant efficiency gains for HQC’s fixed-weight secrets.

\subsubsection{AVX-Aligned Word and Bit Rotation}
Each cyclic shift decomposes into a coarse word rotation and a fine intra-word bit rotation. For word-aligned shifts, operands rotate by entire words and accumulate directly; for others, adjacent words share carry bits to maintain continuity across 64-bit boundaries. Rotated results are XORed in parallel across vector lanes.  

Polynomial buffers are stored as contiguous word arrays aligned to AVX2 widths, enabling uniform memory access. A final masking step enforces the $(X^N-1)$ modulus when the polynomial length is not a multiple of the word size. This layout sustains high throughput, minimizes control divergence, and preserves constant-time execution independent of secret data.

Such alignment and rotation are essential for accelerating dense-by-sparse polynomial multiplication, which is one of the dominant operations in HQC.

\subsubsection{Threshold-Aware Scheduling}
A hybrid scheduler selects between sparse–dense and dense kernels based on the observed Hamming weight. For small fixed weights $\omega$, the sparse path replaces convolution with $\omega$ cyclic shifts and XORs; for denser inputs, it falls back to the dense routine with modular reduction.  

A tunable sparsity threshold, calibrated per parameter set, captures cache and vector-width effects. When placed near the empirical cross-over point, the sparse path removes large constant factors while the fallback path matches dense performance with no regression. The multiplication stage thus becomes bandwidth-bound and highly vectorizable, complementing the system’s AVX2 design choices to reduce end-to-end KEM latency.

\subsection{AVX-Accelerated SHAKE (Keccak)}
\label{sec:shake}

SHAKE contributes roughly $5\%$ of end-to-end HQC cost, mainly from seed expansion, noise sampling, and key derivation. We target this by widening the Keccak’s I/O and exploiting inter-message parallelism.

\subsubsection{Vectorized I/O and round staging}
The Keccak state and message buffers are arranged for 256-bit loads/stores during absorb and squeeze, reducing loop iterations and consolidating data movement. The permutation ($\theta,\rho,\pi,\chi,\iota$) remains constant-time in 64-bit operations, while state transfers occur in wide, fixed-stride bursts to keep the vector unit busy and avoid data-dependent control flow.

\subsubsection{Multi-buffer parallelism and integration}
When multiple SHAKE instances arise naturally (e.g., in \textit{KeyGen}, \textit{Encrypt}, \textit{Decrypt}), they are processed in small, cache-friendly batches. This preserves interfaces and side-channel properties while measurably reducing SHAKE time within the overall pipeline.

\begin{figure*}[t]
    \centering
    \includegraphics[width=0.87\linewidth]{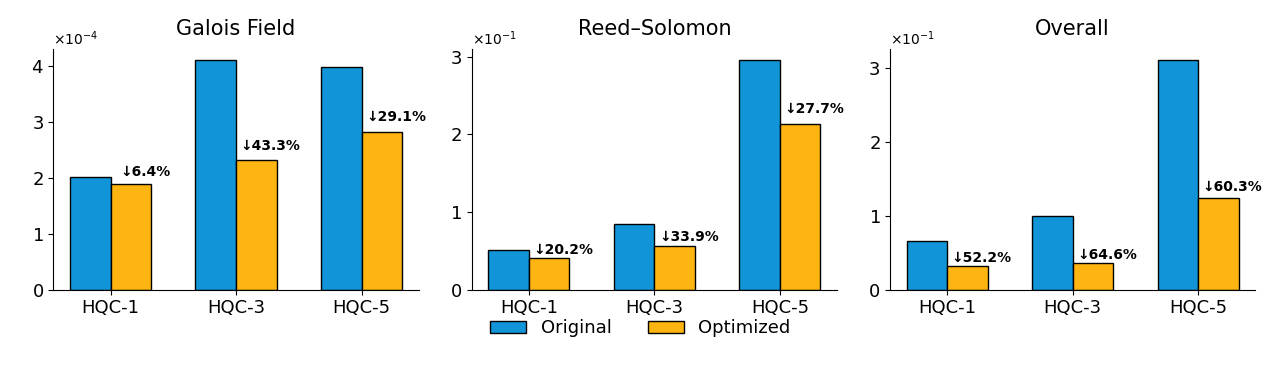}
    \vspace{-0.4cm}
    \caption{Runtime of HQC component, original \cite{hqc_gitlab} VS. our OptHQC. The lower the better. The overall graph shows the overall runtime improvement when we combine all our implemented optimization methods.}
    \label{fig:hqc_perf_bar}
    \vspace{-0.5cm}
\end{figure*}

\subsection{Table-driven Reed--Solomon (Encode) and Syndrome (Decode)}
\label{sec:enc}
\begin{figure}[t]
    \centering
    \includegraphics[width=0.45\textwidth]{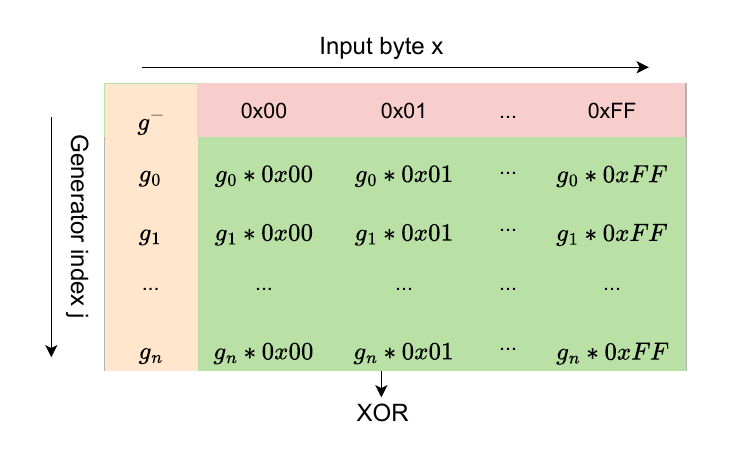}
    \vspace{-0.1cm}
    \caption{
        Lookup table optimization for Reed--Solomon codes. 
        Each entry stores $g_j \cdot x \in \mathrm{GF}(2^8)$, 
        replacing runtime multiplications with constant-time lookups. }
    \label{fig:rs_lookup_table}
    \vspace{-0.6cm}
\end{figure}

The Reed--Solomon encoding process transforms a message into a longer sequence (codeword) through multiple stages of polynomial multiplication. This procedure is computationally intensive, as the polynomial operations are interdependent and thus difficult to parallelize effectively. Moreover, since these multiplications occur based on the number of symbols ($k$), they are not constant-time operations, potentially introducing timing side channels that could leak information.

We propose to replace per-byte multiplications with a precomputed lookup table, as illustrated in Figure~\ref{fig:rs_lookup_table}. Specifically, a static array (approximately 8\,KB) is initialized on first use, storing the products for every combination of generator coefficient $g_j$ and byte value $x$.
The streaming encoder maps each data byte directly to the lookup table, leaving the standard parity shift-register logic intact. This transformation reduces the inner loop to pure XOR and shift operations with linear, cache-friendly memory access. It makes each byte access a constant-time table lookup while preserving deterministic behavior and side-channel resistance.

On the decode path, we apply the same optimization technique to significantly reduce the cost of compute syndromes using two nibble-sliced lookup tables (approximately 14\,KB in total). A lazy initializer precomputes the contributions of $\alpha^{i(j-1)}$ for all high and low nibbles, allowing each byte multiplication to be resolved through a pair of XOR lookups (\emph{hi} $\oplus$ \emph{lo}) without branching. The routine remains constant-time and streaming-compatible while eliminating roughly 1{,}350 $\mathrm{GF}(2^8)$ multiplications per codeword.

\section{Experimental Results}

\subsection{Experiment Setting}

All experiments were conducted on a Dell Optiplex 5050 desktop equipped with an Intel Core i7-7700 CPU 4 cores, 8 threads, 3.6\,GHz, 16\,GB DDR4-2400\,MHz RAM, running Ubuntu 20.04 LTS. 
The HQC reference implementation was obtained from the official HQC website \cite{hqc_gitlab} and compared against our optimized \textbf{OptHQC} framework when running on our platform. %Compilation was performed using GCC 11.3.

All three standardized parameter sets (HQC-1, HQC-3, and HQC-5) were benchmarked across the three cryptographic operations: \textit{KeyGen}, \textit{Encrypt}, and \textit{Decrypt}. 
Each operation was executed repeatedly to mitigate system jitter, and average timings were recorded using both the \texttt{rdtsc} cycle counter and high-resolution wall-clock timing. 
Functional correctness was verified against the official test vectors through the built-in unit test suite to ensure bit-for-bit consistency of ciphertexts and shared secrets. 
All optimizations were compiled in constant-time mode to preserve side-channel resistance.

\subsection{Results and Analysis}

Figure~\ref{fig:hqc_perf_bar} summarizes the per-component runtime of the original and optimized HQC implementations, and the overall improvement of the algorithm. 
Across all parameter sets, the optimized design achieves 50-60\% reduction in total runtime, with the largest improvements observed in \textit{Decryption} due to the enhanced Reed--Solomon and Reed--Muller decoding pipeline. 

Performance improvements remain consistent across all security levels, indicating that the optimizations scale proportionally with code length and parameter size rather than relying on any specific configuration. 
Notably, HQC-3 benefits the most due to its higher decoding depth, which amplifies the efficiency of the optimized Reed--Solomon component.

\paragraph*{Component-Level Acceleration}
Profiling results reveal that \textit{polynomial multiplication} and \textit{concatenated decoding} dominate HQC’s runtime, jointly contributing over 85\% of total latency. 
The optimized implementation integrates three key accelerations:
\begin{itemize}
    \item \textbf{Sparse $\times$ Dense AVX2 Multiplication:} A weight-aware rotation and shift--XOR kernel replaces dense convolution, yielding 20--40\% faster multiplication depending on sparsity.
    \item \textbf{AVX-Parallel SHAKE:} Lane-interleaved Keccak processing accelerates seed expansion by \textbf{2$\times$}, reducing KeyGen overhead.
    \item \textbf{Table-Driven Reed--Solomon Decoder:} Precomputed GF($2^8$) products convert inner-loop multiplications into cache-friendly XOR lookups, reducing \textit{GF\_mul} runtime by 20--25\%.
\end{itemize}

\section{Conclusion}
This work presents a performance analysis and optimization framework for the HQC post-quantum key encapsulation mechanism.
By profiling its main computation blocks, such as hashing, sampling, encoding, syndrome computation, and concatenated Reed–Muller/Reed–Solomon decoding, we identified bottlenecks and applied targeted optimizations. An optimized sparse-by-dense polynomial multiplication using vectorized shifting achieves up to 40\% faster encryption, while a table-driven Reed–Solomon encoder/decoder replaces costly field multiplications with constant-time, cache-friendly operations. These optimizations collectively enhance HQC efficiency and side-channel resistance, demonstrating the potential of memory-efficient design for accelerating code-based PQC on general-purpose platforms.

\clearpage
\bibliographystyle{IEEEtran.bst}
\bibliography{reference}

% \bibliographystyle{ACM-Reference-Format.bst}
% \bibliography{reference.bib}
%\balance
\end{document}